\documentclass[a4paper]{article}
\usepackage{mathrsfs}
\usepackage{fullpage,amsmath,amsfonts,amsthm}
\usepackage{amssymb}
\usepackage{cite}
\newcommand\sA{\mathcal A}
\renewcommand\tilde\widetilde
\renewcommand\hat\widehat

\theoremstyle{example}

\theoremstyle{definition}

\newtheorem{rem}{Remark}

\theoremstyle{plain}

\newtheorem{thm}{Theorem}

\begin{document}
    \title{Quasideterminant solutions to a noncommutative q-difference two-dimensional Toda lattice equation}
    \author{C.X. Li$^{1,2}$\footnote{trisha\_li2001@163.com}, J.J.C. Nimmo$^{3}$\footnote{j.nimmo@maths.gla.ac.uk} and Shoufeng Shen$^{4,2}$\\$^{1}$School of Mathematical Sciences,\\
    Capital Normal University,
    Beijing 100048, CHINA\\
    $^{2}$ Department of Mathematics and Statistics,\\
    University of South Florida,
Tampa, FL 33620-5700, USA\\
     $^{3}$Department of Mathematics,\\
    University of Glasgow,
    Glasgow G12 8QW, UK,\\
    $^{4}$Department of Applied Mathematics,\\
    Zhejiang University of Technology, Hangzhou 310023, CHINA
    }
    \date{}
\maketitle

\begin{abstract}
In \cite{LN}, a generalized type of Darboux transformations defined in terms of a twisted derivation was constructed in a unified form. Such twisted derivations include regular derivations, difference operators, superderivatives and $q$-difference operators as special cases. The formulae for the iteration of Darboux transformations are expressed in terms of quasideterminants.  This approach not only enables one to recover the known Darboux transformations and quasideterminant solutions to the noncommutative KP equation, the non-Abelian two-dimensional Toda lattice equation, the non-Abelian Hirota-Miwa equation and the super KdV equation, but also inspires us to investigate quasideterminant solutions to q-difference soliton equations. In this paper, we first derive the bilinear B\"acklund transformations for the known bilinear $q$-difference two-dimensional Toda lattice equation (q-$2$DTL), then derive a Lax pair whose compatibility gives a new nonlinear q-$2$DTL equation and finally obtain its quasideterminant solutions by iteration of its Darboux transformations.
\end{abstract}

\section{Introduction}
Recently, there has been great interest in noncommutative (nc) versions of soliton equations and their integrability. Examples include the KP equation, the modified KP equation, the KdV equation, the Hirota-Miwa equation and the two-dimensional Toda lattice equation\cite{K,P,S,WW1,WW2,WW3,H,HT1,DH,JN,LN1,CGJN,GN2,LN2,LN3}. Throughout the paper, we do not assume that in general, the dependent variables commute. Results obtained under this assumption are applicable to most of the nc integrable systems. It has been shown that nc integrable systems often have solutions expressed in terms of quasideterminants\cite{GR}. For instance, in \cite{CGJN}, two families of solutions of the nc KP equation were presented which were termed quasiwronskians and quasigrammians. The origin of these solutions was explained by Darboux and binary Darboux transformations. The quasideterminant solutions were then verified directly using formulae for derivatives of quasideterminants (see also \cite{DH}).

Supersymmetric equations can be looked on as a special type of nc integrable systems due to the presence of odd variables. In \cite{LN}, we proposed a twisted derivation following the terminology used in \cite{HLS,CDC,DAMF}. This twisted derivation includes regular derivations, difference operators, superderivatives and $q$-difference operators as some of its special cases. We showed that one can formulate Darboux transformations for such twisted derivations and the iteration formulae are expressed in terms of quasideterminants in which one simply replaces the derivative with the twisted derivation. As an example, we obtained solutions to the super KdV equation expressed in terms of quasideterminants in a unified way for the first time. In fact, the super KdV equation has solutions expressed in terms of superdeterminants \cite{LM} and the relations between superdeterminants and quasideterminants have been established in the literature \cite{BFA,DWB}. However, in \cite{LM}, the authors had to consider different cases in order to obtain superdeterminant solutions and failed to express solutions in terms of superdeterminants in one case. By using quasideterminants, one can avoid such discussions. The above results not only enable us to recover the Darboux transformations and therefore quasideterminant solutions to the existing nc KP equation, the non-Abelian $(2+1)$-dimensional Toda lattice equation and the non-Abelian Hirota-Miwa equation, but also inspire us to seek quasideterminant solutions to q-difference soliton equations. To a certain point, the proposal of the twisted derivation together with its Darboux transformations provide a systematic approach to study some supersymmetric equations, difference equations, differential equations and $q$-difference soliton equations and their quasideterminant solutions.

The $q$-difference (also called $q$-analogue, $q$-deformed or $q$-discrete) integrable system is defined by means of $q$-difference operator ($q$-derivative or Jackson derivative) instead of the normal derivative $\partial$ with respect to independent variables $x,y$ or $t$, etc. in a classical system \cite{KV}. It reduces to a classical integrable system as $q\rightarrow 1$. The study of $q$-analogues of classical integrable systems in parallel with classical integrable systems has attracted much attention. Examples include $q$-difference Pain\'eve equations, $q$-deformed KdV hierarchy and mKdV hierarchy, $q$-deformed KP hierarchy and constrained KP hierarchy, $q$-discrete two-dimensional Toda molecule ($q$-$2$DTM) equation and $q$-$2$DTL equation were studied as well as their integrability \cite{q-P3,q-P6,q-KdV,q-mKdV,q-KP,KOS}. In \cite{KOS}, a $q$-$2$DTM equation and a $q$-$2$DTL equation as well as their determinant solutions were presented. The B\"acklund transformation and Lax pair were obtained for the former, but the B\"acklund transformation and Lax pair for the latter remains unknown. As another application of the twisted derivation, we will take the $q$-$2$DTL equation which involves $q$-difference operators as an example to explore its quasideterminant solutions.

In this paper, we will first present the bilinear B\"acklund transformation and Lax pair for the $q$-$2$DTL equation and then construct its quasideterminant solutions by iterating its Darboux transformations. The paper is organized as follows. In Section 2 and Section 3, we give a brief review of quasideterminants and Darboux transformations for the twisted derivation respectively and discuss their applications in some known nc integrable systems. In Section 4, we derive B\"acklund transformation and Lax pair for the $q$-$2$DTL equation. We present quasideterminant solutions for the $q$-$2$DTL equation constructed by Darboux transformations in Section 5 and give conclusions in Section 6.

\section{Quasideterminants}
In this short section we will list some of the key elementary
properties of quasideterminants used in the paper. The reader is referred to the
original papers \cite{GR,EGR,GGRL} for a more detailed and general
treatment.

An $n\times n$ matrix $M=(m_{i,j})$ over a ring $\mathcal R$ (noncommutative,
in general) has $n^2$ \emph{quasideterminants} written as
$|M|_{i,j}$ for $i,j=1,\dots, n$, which are also elements of
$\mathcal R$. They are defined recursively by

\begin{align}\label{defn}
    |M|_{i,j}&=m_{i,j}-r_i^j(M^{i,j})^{-1}c_j^i,\quad M^{-1}=(|M|_{j,i}^{-1})_{i,j=1,\dots,n}.
\end{align}
In the above $r_i^j$ represents the $i$th row of $M$ with the $j$th
element removed, $c_j^i$ the $j$th column with the $i$th element
removed and $M^{i,j}$ the submatrix obtained by removing the $i$th
row and the $j$th column from $M$. Quasideterminants can be also
denoted as shown below by boxing the entry about which the expansion
is made
\[
|M|_{i,j}=\begin{vmatrix}
    M^{i,j}&c_j^i\\
    r_i^j&\fbox{$m_{i,j}$}
    \end{vmatrix}.
\]
Note that if the entries in $M$ commute then
\begin{equation}\label{commute}
|M|_{i,j}=(-1)^{i+j}\frac{\det(M)}{\det(M^{i,j})}.
\end{equation}


\paragraph{Noncommutative Jacobi Identity}

There is a quasideterminant version of Jacobi identity for determinants, called the noncommutative Sylvester's Theorem
by Gelfand and Retakh \cite{GR}. The simplest version of this identity is given by
\begin{equation}\label{nc syl}
    \begin{vmatrix}
      A&B&C\\
      D&f&g\\
      E&h&\fbox{$i$}
    \end{vmatrix}=
    \begin{vmatrix}
      A&C\\
      E&\fbox{$i$}
    \end{vmatrix}-
    \begin{vmatrix}
      A&B\\
      E&\fbox{$h$}
    \end{vmatrix}
    \begin{vmatrix}
      A&B\\
      D&\fbox{$f$}
    \end{vmatrix}^{-1}
    \begin{vmatrix}
      A&C\\
      D&\fbox{$g$}
    \end{vmatrix},
\end{equation}
where $f,g,h,i\in\mathcal R$, $A$ is an $n\times n$ matrix and $B,C$ (resp. $D,E$) are column (resp. row) $n$-vectors over $\mathcal R$.
As a direct result, we have
the homological relation
\begin{align}
\begin{vmatrix}
      A&B&C\\
      D&f&\fbox{$g$}\\
      E&h&i
    \end{vmatrix}
&=      \begin{vmatrix}
      A&B&0\\
      D&f&\fbox{0}\\
      E&h&1
    \end{vmatrix}
    \begin{vmatrix}
      A&B&C\\
      D&f&g\\
      E&h&\fbox{$i$}
    \end{vmatrix}.\label{HLR}
\end{align}
\paragraph{Quasi-Pl\"{u}cker coordinates} Given an $(n+k)\times n$
matrix $A$, denote the $i$th row of $A$ by $A_i$, the submatrix of
$A$ having rows with indices in a subset $I$ of $\{1,2,\dots,n+k\}$
by $A_I$ and $A_{\{1,\dots,n+k\}\backslash\{i\}}$ by
$A_{\hat\imath}$. Given $i,j\in\{1,2,\dots,n+k\}$ and $I$ such that
$\#I=n-1$ and $j\notin I$, one defines the \emph{(right)
quasi-Pl\"{u}cker coordinates}
\begin{equation}\label{rplucker}
    r^I_{ij}=r^I_{ij}(A):=
    \begin{vmatrix}
    A_I\\
    A_i
    \end{vmatrix}_{ns}
    \begin{vmatrix}
    A_I\\
    A_j
    \end{vmatrix}_{ns}^{-1}=-
    \begin{vmatrix}
    A_I&0\\
    A_i&\fbox{0}\\
    A_j&1
    \end{vmatrix},
\end{equation}
for any column index $s\in\{1,\dots,n\}$. The final equality in
\eqref{rplucker} comes from an identity of the form \eqref{nc syl}
and proves that the definition is independent of the choice of
$s$.

\section{A type of twisted derivations and Darboux transformations}\label{sec:DT}
In \cite{LN}, we proposed a type of twisted derivations and constructed Darboux transformations in terms of it in a unified way. Such twisted derivations include regular derivatives, difference operators, superderivatives and $q$-difference operators as special cases. Surprisingly, the formulae for the iteration of Darboux transformations were expressed in terms of quasideterminants. So far, we have succeeded in deriving quasideterminant solutions to some nc soliton equations involving regular derivatives, difference operators and superderivatives. These results inspire us to consider quasideterminant solutions to $q$-difference soliton equations. Here let us first give a brief review on some known results.
\subsection{A type of twisted derivations}
Consider a general setting in which $\sA$ is an associative, unital algebra over ring $K$, not necessarily graded. Suppose that there is a homomorphism $\sigma\colon\sA\to\sA$ (i.e.\ for all $\alpha\in K$, $a,b\in\sA$, $\sigma(\alpha a)=\alpha \sigma(a)$, $\sigma(a+b)=\sigma(a)+\sigma(b)$ and $\sigma(ab)=\sigma(a)\sigma(b)$) and a \emph{twisted derivation} or $\sigma$-derivation \cite{HLS,CDC,DAMF} $D\colon\sA\to\sA$ satisfying $D(K)=0$ and $D(ab)=D(a)b+\sigma(a)D(b)$.

Simple examples arise in the case that elements $a\in\sA$ depend on a variable $x$, say.
\begin{description}
	\item[Derivative] Here $D=\partial/\partial x$ satisfies $D(ab)=D(a)b+aD(b)$ and $\sigma$ is the identity mapping.
	\item[Forward difference] The homomorphism is the shift operator $T$, where $T(a(x))=a(x+h)$ and the twisted derivation is
	\[
	\Delta(a(x))=\frac{a(x+h)-a(x)}h,
	\]
satisfying $\Delta(ab)=\Delta(a)b+T(a)\Delta(b)$.
		\item[Superderivative] For $a,b\in\sA$, a superalgebra, the superderivative $D=\partial_{\theta}+\theta\partial_x$ satisfies $D(ab)=D(a)b+\hat{a}D(b)$, where $\hat\ $ is the grade involution.
\item[Jackson derivative] The homomorphism is a $q$-shift operator defined by $S_q(a(x))=a(qx)$ and the twisted derivation is
	\[
	D_q(a(x))=\frac{a(qx)-a(x)}{(q-1)x}.
	\]
satisfying $D_q(ab)=D_q(a)b+S_q(a)D_q(b)$.
\end{description}

In what follows several simple properties of twisted derivations which will be used in detailed calculations later on, are listed .
\begin{enumerate}
\item Given $a, b\in\sA$. Whenever $ab$ is defined, $\sigma(ab)=\sigma(a)\sigma(b)$ and $D(ab)=D(a)b+\sigma(a)D(b)$,
\item Let $a$ be invertible over $\sA$. Then $\sigma(a)^{-1}=\sigma(a^{-1})$ and $D(a^{-1})=-\sigma(a)^{-1}D(a)a^{-1}$,
\item Take $a,b,c\in \sA$ such that $ab^{-1}c$ is well-defined. Then
\[
    D(ab^{-1}c)=D(a)b^{-1}c+\sigma(a)\sigma(b)^{-1}(D(c)-D(b)b^{-1}c).
\]
\end{enumerate}

\subsection{Darboux transformations for twisted derivations}
Let $\theta_0,\theta_1,\theta_2,\ldots$ be a sequence in $\sA$. Consider the sequence $\theta[0],\theta[1],\theta[2],\ldots$ in $\sA$, generated from the first sequence by Darboux transformations of the form
\begin{equation}
	G_\theta=\sigma(\theta) D\theta^{-1}=D-D(\theta)\theta^{-1},
\end{equation}
where $D$ and $\sigma$ are the twisted derivation and homomorphism defined above. To be specific, $\theta[0]=\theta_0$ and $G[0]=G_{\theta[0]}$, then let
\begin{equation}\label{theta[1]}
	\theta[1]=G[0](\theta_1)=D(\theta_1)-D(\theta_0)\theta_0^{-1}\theta_1
\end{equation}
and $G[1]=G_{\theta[1]}$, $\theta[2]=G[1]\circ G[0](\theta_2)$ and $G[2]=G_{\theta[2]}$ and so on. In general, for $k\in\mathbb{N}$,
\begin{equation}\label{G[k]}
\theta[k]=G[k-1]\circ G[k-2]\circ\dots\circ G[0](\theta_k),\quad G[k]=\sigma(\theta[k])D\theta[k]^{-1},	
\end{equation}
and we require that each $\theta[k]$ is invertible.

In the standard case, $D=\partial$ and $\sigma=\mathrm{Id}$, it is well known that the terms in the sequence of Darboux transformations have closed form expressions in terms of the original sequence. In the case that $\sA$ is commutative, they are expressed as ratios of wronskian determinants \cite{Crum},
\begin{equation}
\theta[n]=\frac{
\begin{vmatrix}
\theta_0&\dots&\theta_{n-1}&\theta_n\\
\theta^{(1)}_0&\dots&\theta^{(1)}_{n-1}&\theta^{(1)}_n\\
\vdots&&\vdots\\
\theta^{(n-1)}_0&\dots&\theta^{(n-1)}_{n-1}&\theta^{(n-1)}_{n}\\
\theta^{(n)}_0&\dots&\theta^{(n)}_{n-1}&\theta^{(n)}_n
\end{vmatrix}
}{
\begin{vmatrix}
\theta_0&\dots&\theta_{n-1}\\
\theta^{(1)}_0&\dots&\theta^{(1)}_{n-1}\\
\vdots&&\vdots\\
\theta^{(n-1)}_0&\dots&\theta^{(n-1)}_{n-1}
\end{vmatrix}
},\quad n\in\mathbb{N},	
\end{equation}
where $\theta^{(i)}_j$ denotes $\partial^i(\theta_j)$. In the case that $\sA$ is not commutative, the terms in the sequence are expressed as quasideterminants \cite{Crum},
\begin{equation}
\theta[n]=
\begin{vmatrix}
\theta_0&\dots&\theta_{n-1}&\theta_n\\
\theta^{(1)}_0&\dots&\theta^{(1)}_{n-1}&\theta^{(1)}_n\\
\vdots&&\vdots\\
\theta^{(n-1)}_0&\dots&\theta^{(n-1)}_{n-1}&\theta^{(n-1)}_{n}\\
\theta^{(n)}_0&\dots&\theta^{(n)}_{n-1}&\fbox{$\theta^{(n)}_n$}
\end{vmatrix},\quad n\in\mathbb{N}.
\end{equation}
The following theorem gives a generalisation of this formula to the case of general $D$ and $\sigma$. Note in particular that the expressions do not depend on $\sigma$ and are obtained simply by replacing $\partial$ with $D$.
\begin{thm}\label{thm:DT}
Let $\phi[0]=\phi$ and for $n\in\mathbb{N}$ let
\begin{equation*}
\phi[n]=D(\phi[n-1])-D(\theta[n-1])\theta[n-1]^{-1}\phi[n-1],	
\end{equation*}
where $\theta[n]=\phi[n]|_{\phi\to\theta_n}$.
Then, for $n\in\mathbb{N}$,
\begin{equation}\label{theta[n]}
	\phi[n]=
	\begin{vmatrix}
	    \theta_0&\cdots&\theta_{n-1}&\phi\\
	    D(\theta_0)&\cdots&D(\theta_{n-1})&D(\phi)\\
	    \vdots&&\vdots&\vdots\\
	    D^{n-1}(\theta_0)&\cdots&D^{n-1}(\theta_{n-1})&D^{n-1}(\phi)\\
	    D^{n}(\theta_0)&\cdots&D^{n}(\theta_{n-1})&\fbox{$D^{n}(\phi)$}
	\end{vmatrix}.
\end{equation}
\end{thm}

\subsection{Applications in some known noncommutative integrable systems}
In the literature, quasideterminant solutions to the nc KP equation, the non-Abelian $(2+1)$-dimensional Toda lattice equation, the non-Abelian  Hirota-Miwa equation and the super KdV equation have been constructed by elementary Darboux transformations and binary Darboux transformations separately. In this section, we would like to illustrate how the Darboux transformations for the above-mentioned nc integrable systems coincide with the Darboux transformations in terms of the twisted derivation presented in previous sections. Once we have Darboux transformations, it is not difficult to construct quasideterminant solutions by the standard procedure.
\begin{description}
\item[Example 1:] \textbf{The nc KP equation} (Ref. \cite{CGJN})\\
The nc KP equation reads as
\begin{equation*}
(v_t+v_{xxx}+3v_xv_x)_x+3v_{yy}-3[v_x,v_y]=0.
\end{equation*}
It has the Lax pair
\begin{align*}
L\phi&=(\partial_x^2+v_x-\partial_y)\phi=0,\\
M\phi&=(4\partial_x^3+6v_x\partial_x+3v_{xx}+3v_y+\partial_t)\phi=0.
\end{align*}
Let $\theta$ be a particular eigenfunction of the Lax pair. From Theorem 1, the normal derivative $\partial_x$ induces the Darboux transformation of the general eigenfunction $\phi$ by
\begin{align*}
\tilde\phi=D(\phi)-D(\theta)\theta^{-1}\phi=\phi_x-\theta_x\theta^{-1}\phi
\end{align*}
and further the Darboux transformation of the potential $v$ by
\begin{align*}
\tilde v=v+2\theta_x\theta^{-1}
\end{align*}
which coincide with the results obtained in \cite{CGJN}.
\item[Example 2:] \textbf{The non-Abelian $2$DTL equation} (Ref. \cite{LN2})\\
The non-Abelian $2$DTL equation is written as
\begin{align*}
&U_{n,x}+U_nV_{n+1}-V_nU_n=0,\\
&V_{n,t}+\alpha_nU_{n-1}-U_n\alpha_{n+1}=0.
\end{align*}
It has the Lax pair
\begin{align*}
\phi_{n,x}&=V_n\phi_n+\alpha_n\phi_{n-1},\\
\phi_{n,t}&=U_n\phi_{n+1}.
\end{align*}
Assume $\theta_n$ is a particular eigenfunction of the Lax pair. From Theorem 1, it is not difficult to derive the Darboux transformation of the general eigenfunction $\phi_n$
\begin{align*}
\tilde\phi_n=\phi_{n+1,x}-\theta_{n+1,x}\theta_{n+1}^{-1}\phi_{n+1}=\alpha_{n+1}(\phi_n-\theta_n\theta_{n+1}^{-1}\phi_{n+1}),
\end{align*}
which is almost the same as the result obtained in \cite{LN2}
\begin{align*}
\tilde\phi_n=\phi_n-\theta_n\theta_{n+1}^{-1}\phi_{n+1}.
\end{align*}
Based on this transformation, one can derive the corresponding transformations of $U_n$ and $V_n$ similar to these given by \cite{LN2}.
\item[Example 3:] \textbf{The non-Abelian Hirota-Miwa equation} (Ref. \cite{LN1})\\
The non-Abelian Hirota-Miwa equation is
\begin{align*}
&U_{ij}+U_{jk}+U_{ki}=0,\\
&T_k(U_{ij})+T_i(U_{jk})+T_j(U_{ki})=0,\\
&T_k(U_{ij})U_{ki}=T_j(U_{ki})U_{ij},
\end{align*}
with $i,j,k\in \{1,2,3\}$, $U_{ij}$ etc. depending on discrete variables $n_1,n_2,n_3$ and $T_i$ the shift operator in variable $n_i$ defined by $T_i(X)=X(n_i+1)$.
The non-Abelian Hirota-Miwa equation has the Lax pair
\begin{align*}
a_i^{-1}T_i(\phi)-a_j^{-1}T_j(\phi)=U_{ij}\phi.
\end{align*}
It is not difficult to show that the forward difference operator $\Delta_i=a_i^{-1}(T_i-1)$ is a twisted derivation with $\sigma=T_i$. Thus given a particular eigenfunction $\theta$, we have the Darboux transformation of $\phi$
\begin{align*}
\tilde\phi=\Delta_i(\phi)-\Delta_i(\theta)\theta^{-1}\phi=a_i^{-1}(T_i(\phi)-T_i(\theta)\theta^{-1}\phi)
\end{align*}
and furthermore the Darboux transformation of the potentials
\begin{align*}
\tilde {U}_{ij}=T_j(T_i(\theta)\theta^{-1})U_{ij}\theta T_i(\theta)^{-1},
\end{align*}
which are nothing but the Darboux transformations presented in \cite{LN1}.
\item[Example 4:] \textbf{The super KdV equation} (Ref. \cite{LN})\\
The Manin-Radul super KdV equation is written as
\begin{align*}
\alpha_t&=\frac{1}{4}(\alpha_{xx}+3\alpha D(\alpha)+6\alpha u)_x,\\
u_t&=\frac{1}{4}(u_{xx}+3u^2+3\alpha D(u))_x,
\end{align*}
where $D=\partial_{\theta}+\theta\partial_x$ with $\theta$ an odd Grassmann variable, $u$ and $\alpha$ are even and odd dependent variables respectively.
It has the Lax pair
\begin{align*}
L\phi&=(\partial_x^2+\alpha D+u)\phi=\lambda\phi,\\
\phi_t&=M\phi=(\partial_x^3+\frac{3}{4}((\alpha \partial_x+\partial_x \alpha)D+u\partial_x+\partial_x u))\phi.
\end{align*}
Given an even particular eigenfunction $\theta_0$ of the Lax pair. From Theorem 1, we have the Darboux transformation of eigenfunction $\phi$
\begin{align*}
\tilde\phi&=D(\phi)-D(\theta_0)\theta_0^{-1}\phi,
\end{align*}
and furthermore the Darboux transformations of potentials
\begin{align*}
\tilde\alpha&=-\alpha+2(D(\theta_0)\theta_0^{-1})_x,\\
\tilde u&=u+D(\alpha)-2D(\theta_0)\theta_0^{-1}(\alpha-(D(\theta_0)\theta_0^{-1})_x)\end{align*}
which are exactly the same as the Darboux transformation given in \cite{LN}.
\end{description}

\section{The noncommutative $q$-$2$DTL equation}
 As we know that the twisted derivation includes normal derivative, forward difference operator, super derivative and $q$-difference operator as its special cases. In the previous section, we have already illustrated how to recover Darboux transformations for normal derivative, forward difference operator and super derivative from the Darboux transformation for the twisted derivation by taking some nc integrable systems as examples. All these results inspire us to study nc $q$-difference soliton equations, their Darboux transformations and quasideterminant solutions.

In \cite{KOS}, a $q$-discrete version of the two-dimensional Toda lattice equation was proposed
\begin{align}
\delta_{q^{\alpha},x}V_n(x,y)&=J_n(x,y)V_n(q^{\alpha}x,y)-J_{n+1}(x,y)V_n(x,y),\label{GQST1}\\
\delta_{q^{\beta},y}J_n(x,y)&=V_{n-1}(q^{\alpha}x,y)-q^{\beta}V_{n}(x,q^{\beta}y),\label{GQST2}
\end{align}
where $\delta_{q^{\alpha},x}$ and $\delta_{q^{\beta},y}$ are q-difference operators defined by
\begin{align}
	\delta_{q^{\alpha},x}f(x,y)=\frac{f(x,y)-f(q^{\alpha}x,y)}{(1-q)x},\,\	 \delta_{q^{\beta},y}f(x,y)=\frac{f(x,y)-f(x,q^{\beta}y)}{(1-q)y}.	
\end{align}

Under the dependent variable transformations
\begin{align*}
V_n(x,y)&=\frac{\tau_{n+1}(q^{\alpha }x,y)\tau_{n-1}(x,q^{\beta }y)}{\tau_n(x,q^{\beta }y)\tau_n(x,y)},\\
J_n(x,y)&=\frac{1}{(1-q)x}\left\{\{1+(1-q)^2xy\}\frac{\tau_n(q^{\alpha}x,y)\tau_{n-1}(x,y))}
{\tau_n(x,q^{\beta} y)\tau_{n-1}(q^{\alpha }x,y)}-1\right\},
\end{align*}
the $q$-$2$DTL equations \eqref{GQST1} and \eqref{GQST2} were transformed to the bilinear form
\begin{align}
\label{QTBF}
\{\delta_{q^{\alpha},x}\delta_{q^{\beta},y}\tau_n(x,y)\}\tau_n(x,y)&-\{\delta_{q^{\alpha},x}\tau_n(x,y)\}\{\delta_{q^{\beta},y}\tau_n(x,y)\}\nonumber\\
&=\tau_{n+1}(x,q^{\beta}y)\tau_{n-1}(q^{\alpha}x,y)-\tau_n(q^{\alpha}x,q^{\beta}y)\tau_n(x,y),
\end{align}
whose solutions were given by
\begin{equation}\label{QTBFS}
	\tau_n(x,y)=
	\begin{vmatrix}
	    f_n^{(1)}(x,y)&f_{n+1}^{(1)}(x,y)&\cdots&f_{n+N-1}^{(1)}(x,y)\\
	    f_n^{(2)}(x,y)&f_{n+1}^{(2)}(x,y)&\cdots&f_{n+N-1}^{(2)}(x,y)\\
	    \vdots&\vdots&\cdots&\vdots\\
	    f_n^{(N)}(x,y)&f_{n+1}^{(N)}(x,y)&\cdots&f_{n+N-1}^{(N)}(x,y)
	\end{vmatrix}
\end{equation}
where $f_n^{(k)}, k=1,\cdots,N,$ satisfy the dispersion relations
\begin{equation}\label{DS}
\delta_{q^{\alpha},x}f_n^{(k)}(x,y)=-f_{n+1}^{(k)}(x,y),\,\ \delta_{q^{\beta},y}f_n^{(k)}(x,y)=f_{n-1}^{(k)}(x,y).
\end{equation}

In the following, we will first present a bilinear B\"acklund transformation for \eqref{QTBF} and then derive a Lax pair whose compatibility condition yields a much simpler $q$-$2$DTL equation.

\subsection{Bilinear B\"acklund transformations for \eqref{QTBF}}
Based on the solutions given by \eqref{QTBFS} and determinant identities, we propose the following bilinear B\"acklund transformation
\begin{align}
&\delta_{q^{\beta},y}\tau_n'(x,y)\cdot\tau_n(x,y)-\delta_{q^{\beta},y}\tau_n(x,y)\cdot\tau_n'(x,y)=\tau_{n-1}'(x,y)\tau_{n+1}(x,q^{\beta}y),\label{BBT1}\\
&\delta_{q^{\alpha},x}\tau_n(x,y)\cdot\tau_{n-1}'(x,y)-\delta_{q^{\alpha},x}\tau_{n-1}'(x,y)\cdot\tau_n(x,y)=\tau_{n-1}(q^{\alpha}x,y)\tau_{n}'(x,y),\label{BBT2}
\end{align}
which transforms the solution $\tau_n(x,y)$ given by \eqref{QTBFS} to another solution $\tau_n'(x,y)$ given by
\begin{equation}\label{ns}
	\tau_n'(x,y)=
	\begin{vmatrix}
	    f_n^{(1)}(x,y)&f_{n+1}^{(1)}(x,y)&\cdots&f_{n+N}^{(1)}(x,y)\\
	    f_n^{(2)}(x,y)&f_{n+1}^{(2)}(x,y)&\cdots&f_{n+N}^{(2)}(x,y)\\
	    \vdots&\vdots&\cdots&\vdots\\
	    f_n^{(N+1)}(x,y)&f_{n+1}^{(N+1)}(x,y)&\cdots&f_{n+N}^{(N+1)}(x,y)
	\end{vmatrix}.
\end{equation}

In what follows, we will prove how the bilinear B\"acklund transformation works for given $\tau_n(x,y)$ and $\tau_n'(x,y)$. For the sake of simplicity, we adopt the notations used in \cite{KOS}. We denote
\begin{equation}
\tau_n'(x,y)=|0,1,\dots,N-1,N|
\end{equation}
where ``$j$" is a column vector given by
\begin{equation}
``j"=\begin{pmatrix}
f_{n+j}^{(1)}(x,y)\\
f_{n+j}^{(2)}(x,y)\\
\vdots\\
f_{n+j}^{(N)}(x,y)\\
f_{n+j}^{(N+1)}(x,y)
\end{pmatrix}.
\end{equation}
Similarly, we can denote
\begin{equation}
\tau_n(x,y)=|0,1,\dots,N-1,t|,
\end{equation}
where $t=(0,0,\dots,0,1)^T$ is an $(N+1)\times 1$ matrix.

By virtue of the dispersion relations \eqref{DS}, we have
\begin{align}
&\tau_{n+1}(x,y)=|1,2,\dots,N,t|,\,\ \tau_{n-1}(x,y)=|-1,0,\dots,N-2,t|,\\
&\tau_{n-1}'(x,y)=|-1,0,\dots,N-2,N-1|,\,\ \tau_n(x,q^{-\beta}y)=|0_{q^{-\beta}y},1,\dots,N-1,t|, \\
&\tau_n'(x,q^{-\beta}y)=|0_{q^{-\beta}y},1,\dots,N-1,N|,\,\ \tau_{n-1}'(x,q^{-\beta}y)=|-1_{q^{-\beta}y},0,\dots,N-2,N-1|,\\
&\tau_{n-1}'(q^{-\alpha}x,y)=|-1,0,\dots,N-2,N-1_{q^{-\alpha}x}|,\,\ \tau_n(q^{-\alpha}x,y)=|0,\dots,N-2,N-1_{q^{-\alpha}x},t|,\\
&(1-q)q^{-\beta}y\tau_{n-1}'(x,q^{-\beta}y)=|0_{q^{-\beta}y},0,\dots,N-2,N-1|,\\
&-(1-q)q^{-\alpha}x\tau_n'(q^{-\alpha}x,y)=|0,1,\dots,N-1,N-1_{q^{-\alpha}x}|.
\end{align}
It is not difficult to prove that the Laplacian expansion of the determinant identity
\begin{equation}
\begin{vmatrix}
0_{q^{-\beta}y}&0&1\,\cdots\, N-1&\emptyset&N&t\\
\cdots\cdots\cdots&\cdots\cdots\cdots&\cdots\cdots\cdots&\cdots\cdots\cdots&\cdots\cdots\cdots&\cdots\cdots\cdots\\
0_{q^{-\beta}y}&0&\emptyset&1\,\cdots\, N-1&N&t
\end{vmatrix}\equiv0
\end{equation}
leads to the following equation
\begin{equation}
(1-q)q^{-\beta}y\tau_{n-1}'(x,q^{-\beta}y)\tau_{n+1}(x,y)-\tau_n'(x,q^{-\beta}y)\tau_n(x,y)+\tau_n(x,q^{-\beta}y)\tau_n'(x,y)=0
\end{equation}
which is nothing but \eqref{BBT1} with $y$ replaced by $q^{\beta}y$. Similarly, the following determinant identity
\begin{equation}
\begin{vmatrix}
-1&0\,\cdots\, N-2&\emptyset&N-1&N-1_{q^{-\alpha}x}&t\\
\cdots\cdots\cdots&\cdots\cdots\cdots&\cdots\cdots\cdots&\cdots\cdots\cdots&\cdots\cdots\cdots&\cdots\cdots\cdots\\
-1&\emptyset&0\,\cdots\, N-2&N-1&N-1_{q^{-\alpha}x}&t
\end{vmatrix}\equiv0
\end{equation}
leads to
\begin{equation}
\tau_{n-1}'(x,y)\tau_n(q^{-\alpha}x,y)-\tau_{n-1}'(q^{-\alpha}x,y)\tau_n(x,y)-(1-q)q^{-\alpha}x\tau_n'(q^{-\alpha}x,y)\tau_{n-1}(x,y)=0,
\end{equation}
which is nothing but \eqref{BBT2} by replacing $x$ with $q^{\alpha}x$.

\subsection{A new nonlinear $q$-$2$DTL equation and its Lax pair}
By introducing the eigenfunction $\phi_{n+1}(x,y)=\tau_n(x,y)/\tau_n'(x,y)$, we can derive the following Lax pair from the bilinear B\"acklund transformation \eqref{BBT1} and \eqref{BBT2}
\begin{align}
&\delta_{q^{\beta},y}\phi_{n+1}(x,y)=V_{n}(x,y)\phi_n(x,y),\label{LP1}\\
&\delta_{q^{\alpha},x}\phi_n(x,y)=-\phi_{n+1}(x,y)+J_n(x,y)\phi_n(x,y),\label{LP2}
\end{align}
where
\begin{align}
&V_n(x,y)=\frac{\tau_{n-1}(x,y)\tau_{n+1}(x,q^{\beta}y)}{\tau_n(x,y)\tau_n(x,q^{\beta}y)},\\
&J_n(x,y)=\frac{1}{(1-q)x}\left(1-\frac{\tau_{n-1}(x,y)\tau_n(q^{\alpha}x,y)}{\tau_n(x,y)\tau_{n-1}(q^{\alpha}x,y)}\right).
\end{align}
The compatibility condition of the Lax pair \eqref{LP1} and \eqref{LP2} gives the following nonlinear $q$-$2$DTL equation
\begin{align}
&\delta_{q^{\beta},y}J_n(x,y)=V_n(x,y)-V_{n-1}(q^{\alpha}x,y),\label{nl1}\\
&\delta_{q^{\alpha},x}V_n(x,y)=J_{n+1}(x,q^{\beta}y)V_n(x,y)-V_n(q^{\alpha}x,y)J_n(x,y).\label{nl2}
\end{align}

Hereafter, we denote the $q$-difference operators $D_1=\delta_{q^{\alpha},x},\,\ D_2=\delta_{q^{\beta},y}$. $\sigma_1$ and $\sigma_2$ are homomorphism satisfying $\sigma_1 (f(x,y))=f(q^{\alpha}x,y)$ and $\sigma_2 (g(x,y))=g(x,q^{\beta }y)$ respectively. With the new notations, \eqref{nl1} and \eqref{nl2} can be rewritten as
\begin{align}
&D_2(J_n)=V_n-\sigma_1(V_{n-1}),\label{nnl1}\\
&D_1(V_n)=\sigma_2(J_{n+1})V_n-\sigma_1(V_n)J_n.\label{nnl2}
\end{align}
It is not difficult to check that both $D_1$ and $D_2$ are twisted derivations with respect to $\sigma_1$ and $\sigma_2$ individually. Noticing the properties of twisted derivations given by Lemma 1 and introducing new variables $X_n$ where
\[V_n=\sigma_2(X_{n+1})X_n^{-1},\,\ J_n=D_1(X_n)X_n^{-1},\]
\eqref{nnl1} and \eqref{nnl2} can be reduced to one single equation
\begin{align}
D_2(D_1(X_n)X_n^{-1})=\sigma_2(X_{n+1})X_n^{-1}-\sigma_1(\sigma_2(X_{n})X_{n-1}^{-1}).\label{ncq2DTL}
\end{align}
From now on, we refer to \eqref{ncq2DTL} as the nc $q$-$2$DTL equation. In what follows, we will concentrate on looking for its quasideterminant solutions constructed by Darboux transformations.

\begin{rem}
In commutative case, one can easily recover the bilinear $q$-$2$DTL equation \eqref{QTBF} from the nc $q$-$2$DTL equation \eqref{ncq2DTL} by letting $X_n=\tau_n/\tau_{n-1}$.
\end{rem}

\section{Quasicasoratian solutions for \eqref{ncq2DTL}}
In this section, we will first derive the Darboux transformations for the $q$-$2$DTL equation and then construct its quasideterminant solutions by Darboux transformations.
\subsection{Darboux transformations}
As was pointed out in Section 3, a $q$-difference operator is a kind of twisted derivations. Therefore, from the general Darboux transformations defined in terms of twisted derivations, we can get the following Darboux transformation for the nc $q$-$2$DTL equation \eqref{ncq2DTL}.
\begin{align}
\tilde{\phi}_n&=\sigma_1(\theta_n)D_1\theta_n^{-1}\phi_n
=(D_1-D_1(\theta_n)\theta_n^{-1})\phi_n=-\phi_{n+1}+\theta_{n+1}\theta_n^{-1}\phi_n,\\
\tilde{J}_n&=J_{n+1}+\sigma_1(\theta_{n+1}\theta_n^{-1})-\theta_{n+2}\theta_{n+1}^{-1}=\{D_1(\theta_{n+1}\theta_n^{-1})+\sigma_1(\theta_{n+1}\theta_n^{-1})J_n\}\theta_n\theta_{n+1}^{-1},\\
\tilde{V}_n&=\sigma_2(\theta_{n+2}\theta_{n+1}^{-1})V_n\theta_n\theta_{n+1}^{-1}=v_{n+1}-D_2(\theta_{n+2}\theta_{n+1}^{-1}),\\
\tilde{X}_n&=-\theta_{n+1}\theta_n^{-1}X_n,
\end{align}
where $\theta_n$ is a particular solution of the Lax pair \eqref{LP1} and \eqref{LP2}.

\subsection{Quasicasoratian solutions}
Let $\theta_{n,i},i=1,\dots,N$ be a particular set of eigenfunctions of the Lax pair \eqref{LP1} and \eqref{LP2} and introduce the nontation $\Theta_n=(\theta_{n,1},\dots,\theta_{n,N})$. The Darboux transformations for the nc $q$-$2$DTL equation may be iterated by defining
\begin{align}
\phi_n[k+1]&=-\phi_{n+1}[k]+\theta_{n+1}[k]\theta_n^{-1}[k]\phi_n[k],\\
J_n[k+1]&=J_{n+1}[k]+\sigma_1(\theta_{n+1}[k]\theta_n^{-1}[k])-\theta_{n+2}[k]\theta_{n+1}^{-1}[k],\\
V_n[k+1]&=\sigma_2(\theta_{n+2}[k]\theta_{n+1}^{-1}[k])V_n[k]\theta_n[k]\theta_{n+1}^{-1}[k],\\
X_n[k+1]&=-\theta_{n+1}[k]\theta_n^{-1}[k]X_n[k],
\end{align}
where $\phi_n[1]=\phi_n,\, X_n[1]=X_n$ and
\begin{equation}
\theta_n[k]=\phi_n[k]|_{\phi_n\rightarrow\theta_{n,k}}.
\end{equation}

In what follows, we will show by induction that the results of $N$-repeated Darboux transformations $\phi_n[N+1]$ and $X_n[N+1]$ can be expressed as in closed form as quasideterminants
\begin{eqnarray}
&&\phi_n[N+1]=(-1)^N
\begin{vmatrix}
\Theta_{n+N}&\fbox{$\phi_{n+N}$}\\
\Theta_{n+N-1}&\phi_{n+N-1}\\
\vdots&\vdots\\
\Theta_{n}&\phi_{n}
\end{vmatrix},\label{QW1}\\
&&X_n[N+1]=
\begin{vmatrix}
\Theta_{n+N}&\fbox{$0$}\\
\vdots&\vdots\\
\Theta_{n+1}&0\\
\Theta_n&1
\end{vmatrix}X_{n}.\label{QW3}
\end{eqnarray}

The initial case $N=1$ is obviously true. Also using the noncommutative Jacobi identity \eqref{nc syl} and the homological relation \eqref{HLR} we have
\begin{align*}
&\phi_n[N+2]=-\phi_{n+1}[N+1]+\theta_{n+1}[N+1]\theta_n[N+1]^{-1}\phi_{n}[N+1]\\
&=-(-1)^N\begin{vmatrix}
\Theta_{n+N+1}&\fbox{$\phi_{n+N+1}$}\\
\Theta_{n+N}&\phi_{n+N}\\
\vdots&\vdots\\
\Theta_{n+1}&\phi_{n+1}
\end{vmatrix}+(-1)^N\begin{vmatrix}
\Theta_{n+N+1}&\fbox{$\theta_{n+N+1,N+1}$}\\
\Theta_{n+N}&\theta_{n+N,N+1}\\
\vdots&\vdots\\
\Theta_{n+1}&\theta_{n+1,N+1}
\end{vmatrix}\begin{vmatrix}
\Theta_{n+N}&\fbox{$\theta_{n+N,N+1}$}\\
\Theta_{n+N-1}&\theta_{n+N-1,N+1}\\
\vdots&\vdots\\
\Theta_{n}&\theta_{n,N+1}
\end{vmatrix}^{-1}\begin{vmatrix}
\Theta_{n+N}&\fbox{$\phi_{n+N}$}\\
\Theta_{n+N-1}&\phi_{n+N-1}\\
\vdots&\vdots\\
\Theta_{n}&\phi_{n}
\end{vmatrix}\\
&=(-1)^{N+1}\left\{\begin{vmatrix}
\Theta_{n+N+1}&\fbox{$\phi_{n+N+1}$}\\
\Theta_{n+N}&\phi_{n+N}\\
\vdots&\vdots\\
\Theta_{n+1}&\phi_{n+1}
\end{vmatrix}-\begin{vmatrix}
\Theta_{n+N+1}&\fbox{$\theta_{n+N+1,N+1}$}\\
\Theta_{n+N}&\theta_{n+N,N+1}\\
\vdots&\vdots\\
\Theta_{n+1}&\theta_{n+1,N+1}
\end{vmatrix}\begin{vmatrix}
\Theta_{n+N}&\theta_{n+N,N+1}\\
\Theta_{n+N-1}&\theta_{n+N-1,N+1}\\
\vdots&\vdots\\
\Theta_{n}&\fbox{$\theta_{n,N+1}$}
\end{vmatrix}^{-1}\begin{vmatrix}
\Theta_{n+N}&\phi_{n+N}\\
\Theta_{n+N-1}&\phi_{n+N-1}\\
\vdots&\vdots\\
\Theta_{n}&\fbox{$\phi_{n}$}
\end{vmatrix}\right\}\\
&=(-1)^{N+1}\begin{vmatrix}
\Theta_{n+N+1}&\theta_{n+N+1,N+1}&\fbox{$\phi_{n+N+1}$}\\
\Theta_{n+N}&\theta_{n+N,N+1}&\phi_{n+N}\\
\vdots&\vdots&\vdots\\
\Theta_{n+1}&\theta_{n+1,N+1}&\phi_{n+1}\\
\Theta_{n}&\theta_{n,N+1}&\phi_{n}
\end{vmatrix}
\end{align*}
and
\begin{align*}
X_n[N+2]&=\theta_{n+1}[N+1]\theta_n[N+1]^{-1}X_n[N+1]\\
&=\begin{vmatrix}
\Theta_{n+N+1}&\fbox{$\theta_{n+N+1,N+1}$}\\
\Theta_{n+N}&\theta_{n+N,N+1}\\
\vdots&\vdots\\
\Theta_{n+1}&\theta_{n+1,N+1}
\end{vmatrix}\begin{vmatrix}
\Theta_{n+N}&\fbox{$\theta_{n+N,N+1}$}\\
\Theta_{n+N-1}&\theta_{n+N-1,N+1}\\
\vdots&\vdots\\
\Theta_{n}&\theta_{n,N+1}
\end{vmatrix}^{-1}\begin{vmatrix}
\Theta_{n+N}&\fbox{$0$}\\
\vdots&\vdots\\
\Theta_{n+1}&0\\
\Theta_n&1
\end{vmatrix}X_{n}\\
&=\begin{vmatrix}
\Theta_{n+N+1}&\fbox{$\theta_{n+N+1,N+1}$}\\
\Theta_{n+N}&\theta_{n+N,N+1}\\
\vdots&\vdots\\
\Theta_{n+1}&\theta_{n+1,N+1}
\end{vmatrix}\begin{vmatrix}
\Theta_{n+N}&\theta_{n+N,N+1}\\
\Theta_{n+N-1}&\theta_{n+N-1,N+1}\\
\vdots&\vdots\\
\Theta_{n}&\fbox{$\theta_{n,N+1}$}
\end{vmatrix}^{-1}X_{n}\\
\intertext{and then using the quasi-Pl\"ucker coordinate formula \eqref{rplucker}, we get}
&=\begin{vmatrix}
\Theta_{n+N+1}&\theta_{n+N+1,N+1}&\fbox{$0$}\\
\Theta_{n+N}&\theta_{n+N,N+1}&0\\
\vdots&\vdots&\vdots\\
\Theta_{n+1}&\theta_{n+1,N+1}&0\\
\Theta_n&\theta_{n,N+1}&1
\end{vmatrix}X_{n}.
\end{align*}
This proves the inductive step. So far, we have finished proving quasideterminant solutions given by \eqref{QW1} and \eqref{QW3} to the nc q-discrete two-dimensional Toda lattice equation \eqref{ncq2DTL}.
\begin{rem}
In the commutative case, it is clear that \eqref{ncq2DTL} has the vacuum solution $X_n=-1$ and therefore we have $X_n[N]=\tau_n/\tau_{n-1}$ from \eqref{QW3} with $\tau_n$ given by \eqref{QTBFS}.
\end{rem}

\section{Conclusions} In our previous work \cite{LN}, we proposed a twisted derivation which includes normal derivative, forward difference operators, q-difference operators and superderivatives as special cases and constructed its Darboux transformation which has an iteration formula written in terms of a quasideterminant. This result opens the opportunity for an unified approach to Darboux transformations for differential, superderivative, forward difference and $q$-difference operators which makes it possible to construct quasideterminant solutions to noncommutative integrable systems. In this paper we show how this has been achieved for a noncommutative $q$-$2$DTL equation which involves $q$-difference operators besides the nc KP equation, the non-Abelian two-dimensional Toda lattice equation, the non-Abelian Hirota-Miwa equation and the super KdV equation which involve normal derivative, forward difference operators and superderivatives.

To sum up, in this paper, we first derive a bilinear B\"acklund transformation and Lax pair for a nc $q$-$2$DTL equation and then construct quasicasoratian solutions to the nc $q$-$2$DTL equation by its Darboux transformation. But how to construct its binary Darboux transformation and quasigrammian solutions remain unsolved. In some sense, these results extend the applications of quasideterminants in nc integrable systems and extended integrable systems and reveal the advantage of using quasideterminants in dealing with nc integrable systems, q-difference soliton equations and supersymmetric equations.

\section*{Acknowledgement}
This work was supported by the National Natural Science Foundation of China under the grants 11271266 and 11371323 and Beijing Teachers Training Center for Higher Education. This work was finished while the authors Chun-Xia Li and Shou-Feng Shen visiting the University of South Florida. One of the authors, Chun-Xia Li would like to thank Professor Wen-Xiu Ma for his hospitality and valuable discussions.

\end{document}